\documentclass[twocolumn,times]{aastex631}

\shorttitle{Recovery and secular variation of the X-ray polarization in Crab}
\shortauthors{Long et al.}

\graphicspath{{./}{fig/}}

\begin{document}

\title{X-ray polarimetry of the Crab nebula with PolarLight: polarization recovery after the glitch and a secular position angle variation}

\author{Xiangyun Long}
\affiliation{Department of Engineering Physics, Tsinghua University, Beijing 100084, China}

\correspondingauthor{Hua Feng}
\email{hfeng@tsinghua.edu.cn}

\author[0000-0001-7584-6236]{Hua Feng}
\affiliation{Department of Astronomy, Tsinghua University, Beijing 100084, China}
\affiliation{Department of Engineering Physics, Tsinghua University, Beijing 100084, China}

\author{Hong Li}
\affiliation{Department of Astronomy, Tsinghua University, Beijing 100084, China}

\author{Jiahuan Zhu}
\affiliation{Department of Astronomy, Tsinghua University, Beijing 100084, China}

\author{Qiong Wu}
\affiliation{Department of Engineering Physics, Tsinghua University, Beijing 100084, China}

\author{Jiahui Huang}
\affiliation{Department of Engineering Physics, Tsinghua University, Beijing 100084, China}

\author{Massimo Minuti}
\affiliation{INFN-Pisa, Largo B. Pontecorvo 3, 56127 Pisa, Italy}

\author{Weichun Jiang}
\affiliation{Key Laboratory for Particle Astrophysics, Institute of High Energy Physics, Chinese Academy of Sciences, Beijing 100049, China}

\author{Weihua Wang}
\affiliation{Department of Astronomy, School of Physics, Peking University, Beijing 100871, China}

\author{Renxin Xu}
\affiliation{Department of Astronomy, School of Physics, Peking University, Beijing 100871, China}

\author{Enrico Costa}
\affiliation{IAPS/INAF, Via Fosso del Cavaliere 100, 00133 Rome, Italy}

\author{Dongxin Yang}
\affiliation{Department of Engineering Physics, Tsinghua University, Beijing 100084, China}

\author{Saverio Citraro}
\affiliation{INFN-Pisa, Largo B. Pontecorvo 3, 56127 Pisa, Italy}

\author{Hikmat Nasimi}
\affiliation{INFN-Pisa, Largo B. Pontecorvo 3, 56127 Pisa, Italy}

\author{Jiandong Yu}
\affiliation{School of Electronic and Information Engineering,  Ningbo University of Technology, Ningbo, Zhejiang 315211, China}

\author{Ge Jin}
\affiliation{North Night Vision Technology Co., Ltd., Nanjing 211106, China}

\author{Ming Zeng}
\affiliation{Department of Engineering Physics, Tsinghua University, Beijing 100084, China}

\author{Peng An}
\affiliation{School of Electronic and Information Engineering,  Ningbo University of Technology, Ningbo, Zhejiang 315211, China}

\author{Luca Baldini}
\affiliation{INFN-Pisa, Largo B. Pontecorvo 3, 56127 Pisa, Italy}

\author{Ronaldo Bellazzini}
\affiliation{INFN-Pisa, Largo B. Pontecorvo 3, 56127 Pisa, Italy}

\author{Alessandro Brez}
\affiliation{INFN-Pisa, Largo B. Pontecorvo 3, 56127 Pisa, Italy}

\author{Luca Latronico}
\affiliation{INFN, Sezione di Torino, Via Pietro Giuria 1, I-10125 Torino, Italy}

\author{Carmelo Sgr\`{o}}
\affiliation{INFN-Pisa, Largo B. Pontecorvo 3, 56127 Pisa, Italy}

\author{Gloria Spandre}
\affiliation{INFN-Pisa, Largo B. Pontecorvo 3, 56127 Pisa, Italy}

\author{Michele Pinchera}
\affiliation{INFN-Pisa, Largo B. Pontecorvo 3, 56127 Pisa, Italy}

\author{Fabio Muleri}
\affiliation{IAPS/INAF, Via Fosso del Cavaliere 100, 00133 Rome, Italy}

\author{Paolo Soffitta}
\affiliation{IAPS/INAF, Via Fosso del Cavaliere 100, 00133 Rome, Italy}

% 250 words 
\begin{abstract}
We report follow-up observations of the Crab nebula with the PolarLight X-ray polarimeter, which revealed a possible variation in polarization associated with a pulsar glitch in 2019. The new observations confirm that the polarization has recovered roughly 100 days after the glitch. With the new observations, we find that the polarization angle (PA) measured with PolarLight from the total nebular emission has a difference of $18\fdg0 \pm 4\fdg6$ from that measured 42 years ago with OSO-8, indicating a secular evolution of polarization with either the Crab nebula or pulsar.  The long-term variation in PA could be a result of multiple glitches in the history,  magnetic reconnection or movement of synchrotron emitting structures in the nebula, or secular evolution of the pulsar magnetic geometry. 
\end{abstract}

\keywords{Polarimetry (1278) --- Rotation powered pulsars (1408) --- X-ray detectors (1815) --- X-ray sources (1822)}

\section{Introduction}

The Crab nebula is so far the only astrophysical source with a significant polarimetric measurement in the keV band.  In the 1970s, the Bragg polarimeter onboard OSO-8 measured a polarization fraction (PF) of $0.157 \pm 0.015$ with a polarization angle (PA) of $161\fdg1 \pm 2\fdg8$ from the total nebular emission \citep{Weisskopf1976}; a phase-resolved analysis indicates that the pure nebular emission has a PF of $0.192 \pm 0.010$ and a PA of $156\fdg4 + 1\fdg4$ \citep{Weisskopf1978a}.  The above results were obtained in a narrow band around 2.6 keV due to the nature of Bragg diffraction, and consistent results were seen around 5.2 keV (the second order diffraction) with slightly larger uncertainties. 

In 2019, the PolarLight instrument re-detected the polarization from the Crab nebula in the energy band of 3.0--4.5 keV, with PF and PA values consistent with those obtained with OSO-8 within 3$\sigma$ error bounds.  Also, PolarLight revealed a possible variation in polarization (a sudden decrease in PF) coincident in time with the glitch of the Crab pulsar on July 23, 2019. The variation is found to have a significance of 3$\sigma$ using different methods, including the Bayes factors, Bayesian posterior distributions, and bootstrap analysis. This may suggest that the pulsar magnetosphere altered after the glitch, yet to be confirmed by missions in the near future \citep{Weisskopf2016,Zhang2019}.  

The above results are obtained with PolarLight observations spanning over a time from 145 days before the glitch to 132 days after the glitch. The PF seems to have recovered roughly 100 days after the glitch. In order to investigate the source behavior on a longer timescale, follow-up observations with PolarLight were executed.  More observations may also help constrain whether such a variation is indeed associated with pulsar glitches. In this Letter, we update the results with the addition of follow-up observations. 

\section{Observations and analysis}

The observations reported in \citet{Feng2020a} have a total exposure of about 660 ks, executed from March 2019 to December 2019.  Since then, we continue to monitor the source until the end of August 2020, with a gap from mid-May to mid-July due to Sun avoidance. The satellite flies in a Sun synchronous orbit. Observations are performed only at low latitudes where the instrument is free of regions filled with high flux of trapped charged particles. The effective exposure is about 15 minutes in each orbit, or around 150 minutes totally in a day \citep{Li2021}.  In sum,  we have obtained a total exposure of 742~ks in the follow-up observations, covering a span of 244 days. Following \citet{Feng2020a}, we adopt the energy band of 3--4.5 keV, where the polarization measurement is the most sensitive. 

When the source is occulted by the Earth, the instrument collects data from the background, which is mainly due to charged particles in the orbit \citep{Huang2021}.  As the detector is an ionization chamber, it can distinguish part of the background events from source events based on the track images; \citet{Huang2021} also find that a fraction of background (e.g., 28\% in 2--8 keV) is not removable because their energy deposits are in a physical process identical to the detection of source photons.  Instead of using the simple algorithm implemented in \citet{Feng2020a}, here we employ an energy-dependent discrimination technique \citep{Zhu2021} that can remove nearly all of the removable background.   

The methods for data reduction and analysis are the same as those used in \citet{Feng2020a}, and are briefly described here. Readers may refer to \citet{Feng2020a} and references therein for more technical details. Events with 58 pixels or more (after the noise cut) located in the central $\pm7$~mm region are selected for analysis.  After particle discrimination, the background contamination in the energy band of 3--4.5 keV is about 8\% in the on-source observations.  The polarization is calculated using the Stokes parameters, and the intrinsic quantities are inferred with the Bayesian approach. This allows for an unbiased estimate of the PF; the estimate of PA is unbiased due to symmetry.  The errors in this Letter are quoted as the 68\% credible interval from the marginalized posterior distribution at the highest density region.  The mean modulation factor is 0.35 in the above energy band. The polarimetric modulation of the background cannot be constrained (consistent with zero, but with a 90\% upper limit of 22\%) because of low statistics. According to laboratory tests and simulations, plus that the measurements are obtained at different instrument orientations, it is reasonable to assume that the background is unpolarized.  The PF quoted in this work has been corrected for the background. 

We identified an error\footnote{The PA is calculated from $\frac{1}{2} \arctan$ but is incorrectly wrapped into $(0\arcdeg, 180\arcdeg)$ for that point due to a code error. No other numbers or plots in \citet{Feng2020a} are affected by this error.} in the bottom panel of Fig.~2a in \citet{Feng2020a} that the 4th point should be near 130\arcdeg\ instead of 40\arcdeg\ or 220\arcdeg. Due to the large uncertainty on that point, it has no influence to the conclusions at all. We also note that, with the implementation of the new background discrimination technique, the results are well consistent with those reported in \citet{Feng2020a}.

%%%%%%%%%%%%%%%%%%%%%%%%%%%%%%%%%%%%%%%%%%%%%%%%
\begin{figure}[t]
\centering
\includegraphics[width=0.8\columnwidth]{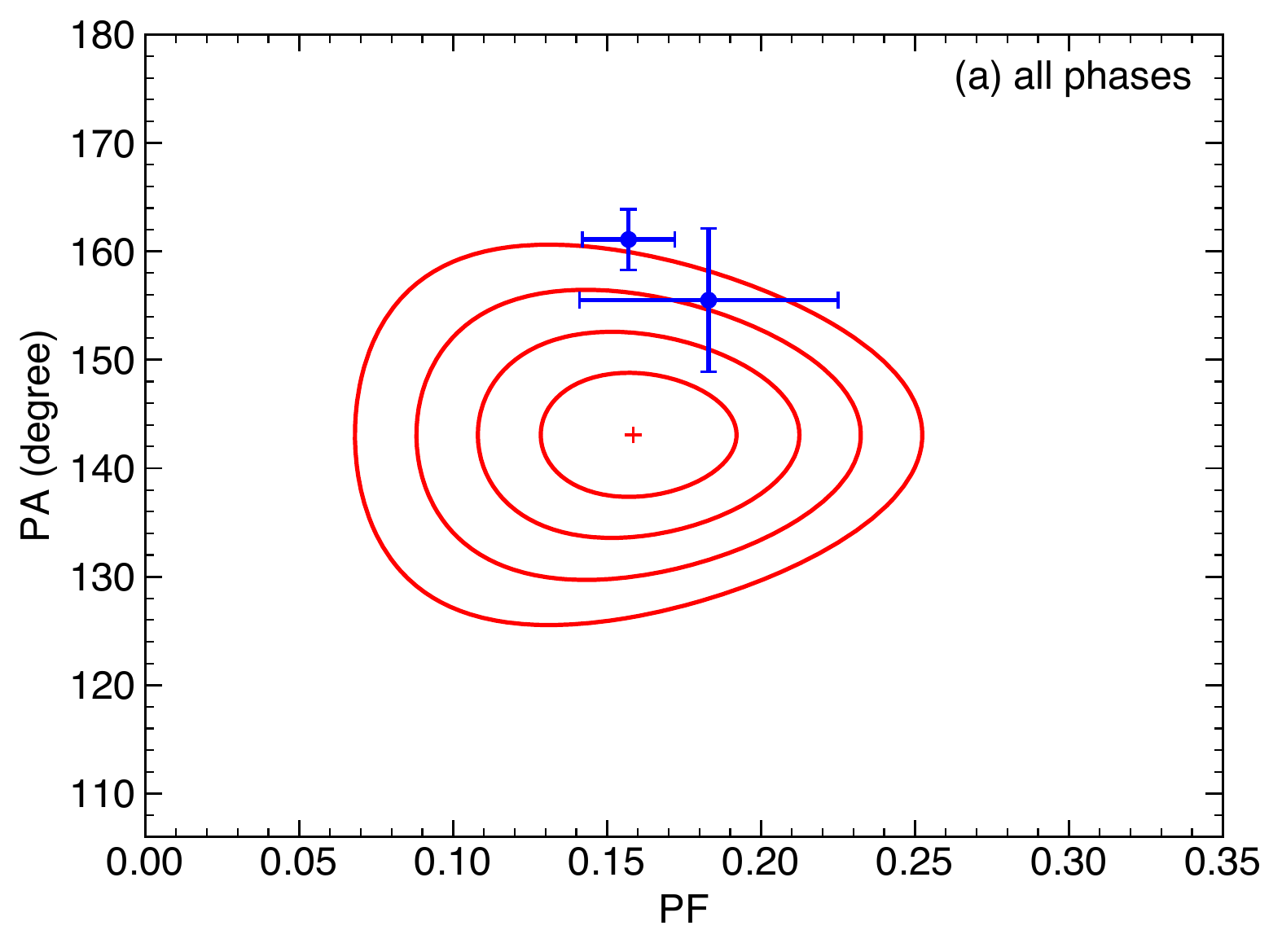}
\includegraphics[width=0.8\columnwidth]{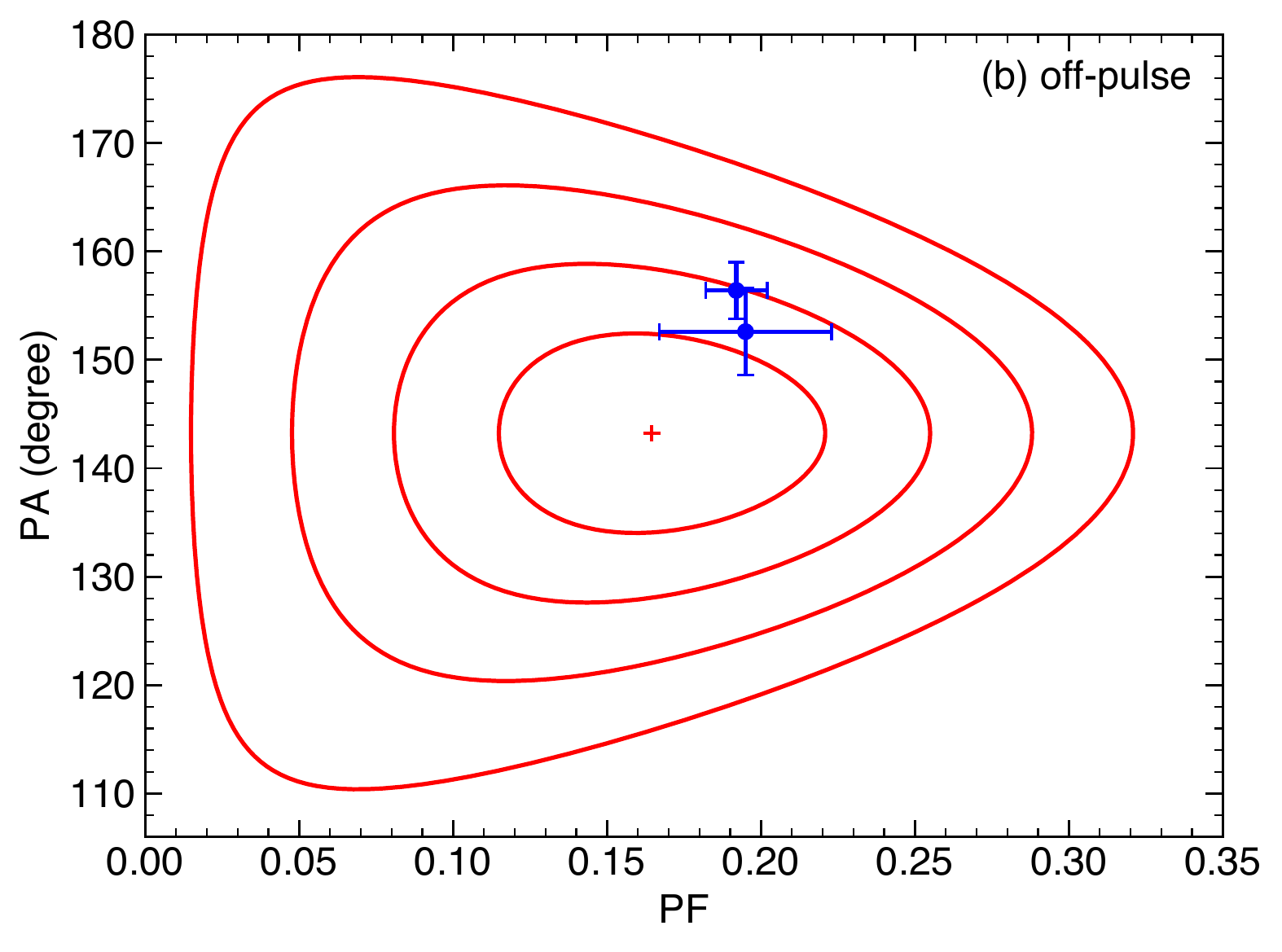}
\caption{X-ray polarization measured with PolarLight in the energy band of 3.0--4.5 keV of the Crab nebula in all pulsar phases (top) and the on-pulse phase interval (bottom). The red plus indicates the point estimate and the contours encircle the 1$\sigma$, 2$\sigma$, 3$\sigma$ and 4$\sigma$ credible intervals. The blue points and error bars mark the measurements with OSO-8 at 2.6 keV (with smaller errors) and 5.2 keV (with larger errors). }
\label{fig:cont}
\end{figure}
%%%%%%%%%%%%%%%%%%%%%%%%%%%%%%%%%%%%%%%%%%%%%%%%

With the full dataset, the polarization is found to have a ${\rm PF} = 0.159_{-0.020}^{+0.021}$ and a ${\rm PA} = 143\fdg1 \pm 3\fdg7$ in all phases, or a ${\rm PF} = 0.165_{-0.035}^{+0.034}$ and a ${\rm PA} = 143\fdg2 \pm 6\fdg0$ in the off-pulse phases.  The definition of the on- and off-pulse phases can be found in \citet{Feng2020a} and is identical to that in \citet{Weisskopf1978a}; the former contains emission from the pulsar while the latter has pure nebular emission.  The results are displayed in Figure~\ref{fig:cont} with 2D confidence contours, in comparison with those measured with OSO-8 \citep{Weisskopf1976,Weisskopf1978a}. Thanks to the improved statistics, the PA in all phases has a smaller error than before but is no longer consistent with that measured with OSO-8. The difference in PA is $18\fdg0 \pm 4\fdg6$, indicative of a secular variation at a significance of 3.9$\sigma$.  Using data before the glitch and those 100 days after the glitch, we have a measurement of ${\rm PA} = 141\fdg2 \pm 3\fdg9$; with data 100 days after the glitch only, we obtain ${\rm PA} = 139\fdg3 \pm 4\fdg8$. Thus, the tension is not alleviated if we exclude the data associated with the possible variation after the glitch. We emphasize that the PA measurement is simply a geometric result and is immune from calibration uncertainties.

\begin{figure}[t]
\centering
\includegraphics[width=0.8\columnwidth]{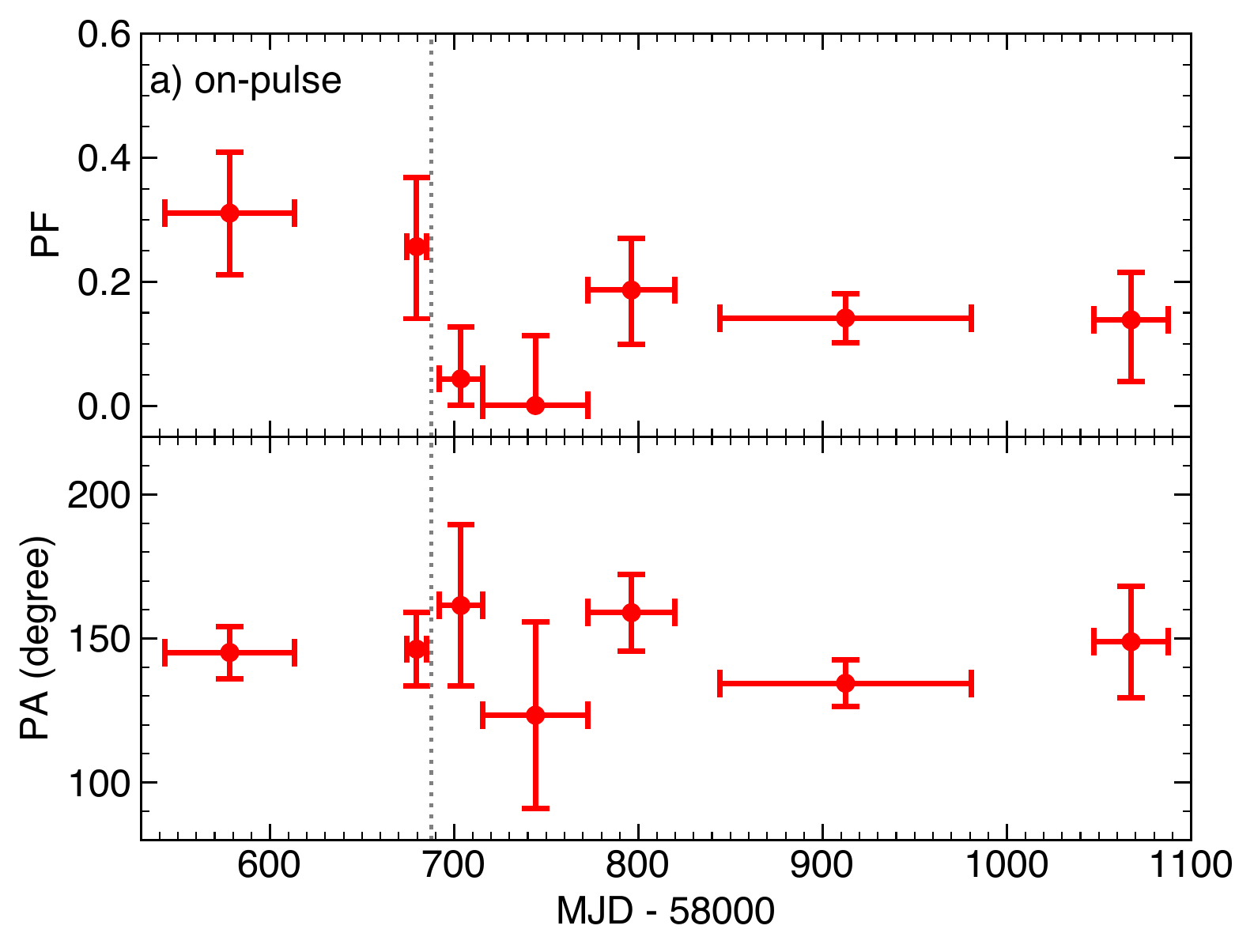}
\includegraphics[width=0.8\columnwidth]{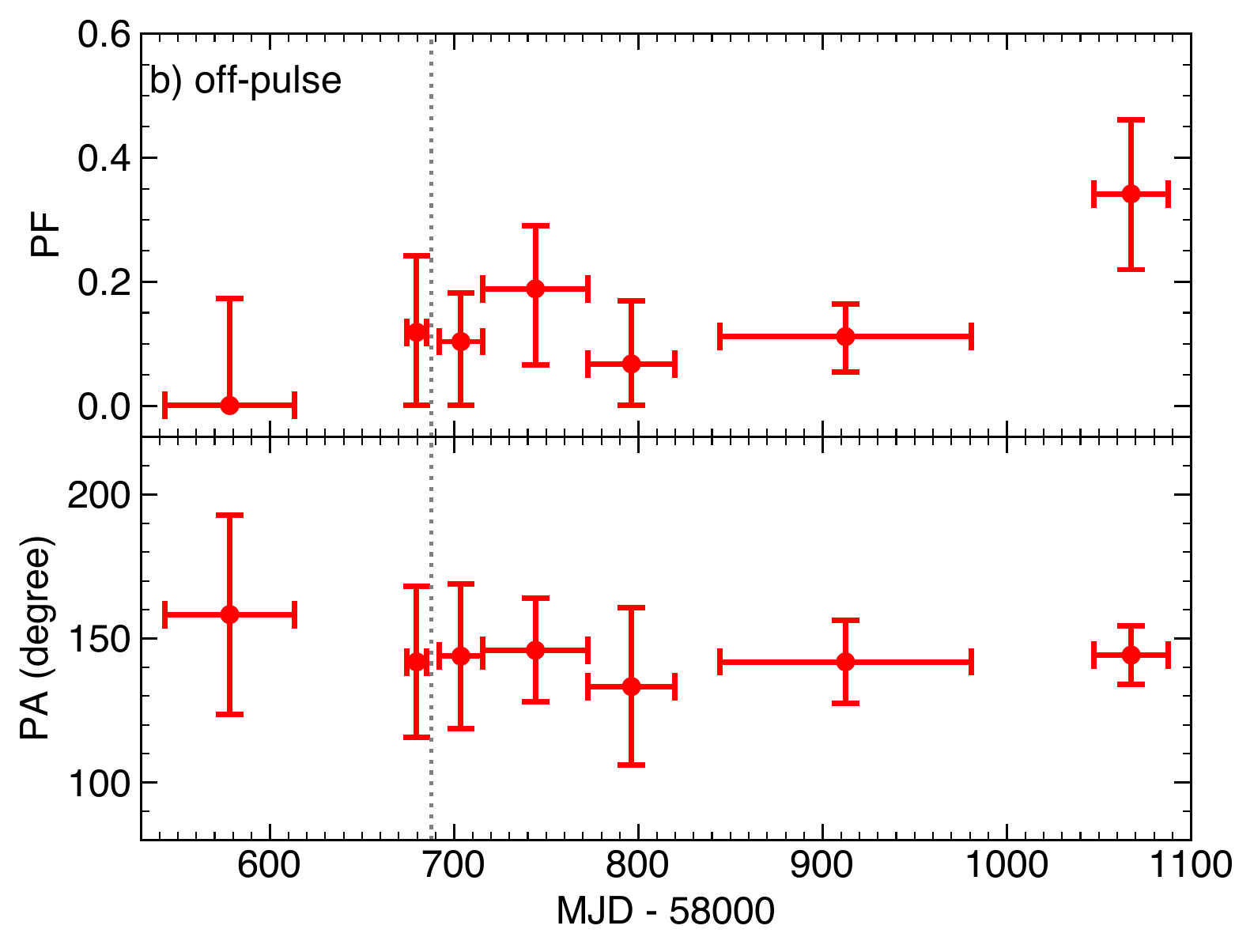}
\caption{The Crab X-ray polarization in the 3.0--4.5 keV band as a function of time, respectively, for events in on-pulse phases (\textbf{a}) and off-pulse phases (\textbf{b}).}
\label{fig:var}
\end{figure}

For the new data (obtained after MJD 58844), we calculate the polarization in two phase intervals, the on-pulse and off-pulse phases, and in two time segments separated by the observing gap to investigate the time variation (Figure~\ref{fig:var}).  Consistent results with larger errors are obtained if the data in the first time segment are further divided into two bins.  In the on-pulse phase, where the variation in polarization was detected, the PF has recovered to a constant level, consistent with the pre-glitch value within errors. The off-pulse polarization is consistent with a constant. 

Inclusion of the the new data does not affect the conclusion in \citet{Feng2020a} that a 3$\sigma$ variation is seen between the data before the glitch and those up to 100 days after the glitch.  The new data confirm that the recovery time is at most 100 days.  Here we employ the Bayes factor method to compare two models based on the full dataset. The first model has a constant PF with time, while the second assumes an instant decrease in PF at the glitch time, followed by an exponential recovery with time.  Compared with the constant model, the second model has two additional parameters, the amplitude of the decrease and the timescale of recovery. Using Monte-Carlo integrations,  the Bayes factor is found to be 0.16, suggesting that the second model is favored with substantial evidence \citep{Jeffreys1961}. 

Without the two data points right after the glitch, the on-pulse PF seems to decline with time (Figure~\ref{fig:var}). We inspect the posterior distributions of the PF and find that they approximate Gaussian distributions. This allows us to use the $\chi^2$ fitting, which suggests that either a constant model or a linear model can produce an acceptable fit, while the linear model is favored at a significance of only 2.2$\sigma$ with F-test. Therefore, there is no evidence for a PF decline. Similarly, there is no statistical evidence for an increase of the off-pulse PF. 

%%%%%%%%%%%%%%%%%%%%%%%%%%%%%%%%%%%%%%%%%%%%%%%%
%%%%%%%%%%%%%%%%%%%%%%%%%%%%%%%%%%%%%%%%%%%%%%%%

\section{Discussion}

In this Letter, we report two findings with the follow-up observations of the Crab nebula with PolarLight. First, we confirm that, over a time span of at most 100 days, the polarization signal has recovered to its pre-glitch level (Figure~\ref{fig:var}).  Radio timing reveals that this glitch has a slow rise followed by an exponential recovery with a time constant of $6.4 \pm 0.4$~d \citep{Shaw2021}, suggesting that the total recovery time is about 30--50 days (5--7 times the time constant). It is shorter but generally consistent with the recovery timescale of the X-ray polarization, perhaps implying a causal connection between the two events.  Due to relatively large errors in polarization measurements, it is unknown whether the recovery in polarization is complete or not. 

The other interesting finding is that we detect a difference of $\Delta{\rm PA} = 18\fdg0 \pm 4\fdg6$ between the PolarLight (from March 2019 to August 2020) and OSO-8 (March 1976 and March 1977) measurements.  The measurements with the two instruments are separated by 42 years, suggesting that there is a secular evolution of polarization.  As the PolarLight measurements cover the occurrence of a glitch, after which a possible variation in polarization is detected, the difference in PA could be a direct result of the glitch.  However, this scenario seems unlikely, because the difference in PA remains if the portion of data most relevant to the post-glitch variation is removed.  On the other hand, there have been 26 glitches including this one detected in the Crab pulsar\footnote{\url{http://www.jb.man.ac.uk/pulsar/glitches.html}} over the past 42 years \citep{Espinoza2011}. If the glitch does trigger a change in the magnetosphere without a full recovery, accumulation of small variations could explain the observed shift in PA.  

In both the optical and gamma-ray bands, emission from the Crab pulsar and a neighboring synchrotron knot shows a large PA variation (25\arcdeg--35\arcdeg) over a course of a few years \citep{Moran2016}.  Also, \citet{Moran2013} revealed a low-significance (2$\sigma$) variation in polarization from the optical knot on a timescale of weeks.  These indicate that the geometry of the magnetic fields changes, which could be a result of magnetic reconnection in the nebula. Thus, a polarimetric variation in the X-ray band from the Crab nebula is not unexpected. 

High resolution imaging observations in the radio, optical, and X-ray bands all revealed a dynamic Crab nebula on timescales from days to months \citep{Hester2002,Bietenholz2004}.  If the magnetic fields have different orientations at different locations, along with the movement of the synchrotron emitting structures in the nebula, one would expect a change in PA.  

Another plausible scenario is that the difference in PA is a result of the long-term evolution of the pulsar magnetosphere. The separation of the two phase peaks, along with other properties in the pulse profile, is found to vary slowly with time, seen in both the radio and X-ray bands \citep{Lyne2013,Ge2016}.  It indicates a secular evolution of the pulsar magnetic geometry, which can explain the detected difference in PA as well. 

\begin{acknowledgments}
We thank the referee for useful comments. HF acknowledges funding support from the National Natural Science Foundation of China under the grant Nos.\ 11633003, 12025301, and 11821303, the CAS Strategic Priority Program on Space Science (grant No.\ XDA15020501-02), and the National Key R\&D Project (grants Nos.\ 2016YFA040080X \& 2018YFA0404502).  
 \end{acknowledgments}
 
 \vspace{5mm}
 \facility{PolarLight}
 
%\bibliography{xpol,unpub,book}{}
%\bibliographystyle{aasjournal}

\end{document}